\documentstyle[12pt]{article}

\begin{document}
\pagestyle{plain}
\centerline{\large {\bf Critical Level Statistics in Two-dimensional}}

\centerline{\large {\bf Disordered Electron Systems}}

\vskip 1cm

\centerline{Tomi O{\footnotesize HTSUKI}\footnote{
Present Address; Department of Physics, Sophia University,
Kioicho 7-1, Tokyo 102}
and Yoshiyuki O{\footnotesize NO}}
\medskip

\centerline{\it Department of Physics, Toho University,}
\centerline{\it Miyama 2--2--1, Funabashi, Chiba 274}

\bigskip

\centerline{(received \hspace{6cm} )}

\vskip 1.5cm

\noindent
{\bf Abstract}{\ }
The level statistics in the two dimensional disordered electron systems
in magnetic fields (unitary ensemble) or
in the presence of strong spin-orbit scattering (symplectic ensemble)
are investigated at the Anderson transition points.
The level spacing distribution functions $P(s)$'s
are found to be independent of the
system size or of the type of the potential distribution, suggesting the
universality.
They behave as $s^2$ in the small $s$ region in the former case,
while $s^4$ rise is seen in the latter.

\vskip 3cm

\noindent
KEYWORDS: level statistics, level spacing distribution, random matrix
theory, Gaussian unitary ensemble, Guassian symplectic ensemble,
Anderson transition, universality

\vfil\eject

\noindent
The random matrix theory, first developed by Wigner in 1950's
in the field of nuclear physics, is now widely applied to many problems
in solid state physics such as the energy spectrum of fine particles,
conductance fluctuations in mesoscopic systems and the persistent current
in mesoscopic rings \cite{hase}.
In metals where the electronic states are  extended,
the level repulsion is strong and the distribution function $P(s)$ for the
level spacing $s$
is well approximated by the Wigner surmise.
On the other hand, in the insulating regime no level repulsion
is expected since the electrons are localized and $P(s)$ becomes
Poissonian.
Natural question that arises from the above facts is how the properties
of the level statistics are changed when the system undergoes the
localization-delocalization transition, i.e., Anderson transition.

To answer this question, much numerical effort has been devoted to
the three dimensional (3D) Anderson model.
Shklovskii {\it et al.} are the first to notice that $P(s)$ at the
Anderson transition becomes size-independent \cite{shkl},
and the critical exponent $\nu$
for the localization length can be extracted from the behavior of the
size dependence
of $P(s)$ near the Anderson transition.
Evangelou {\it et al.} \cite{evan} as well as Hofstetter and his
coworkers \cite{HS}
made detailed analyses on the level statistics at the critical
point, and $P(s)$ is again shown to be size independent and
to be different from the Wigner surmise or Poissonian.

The analytical calculation of $P(s)$ by Kravtsov {\it et al.} \cite{krav}
has suggested the possibility to obtain the critical exponent $\nu$
from the asymptotic behavior of  $P(s)$ at the Anderson transition.
They predicted that $P(s)$ falls off as
\begin{equation}
P(s)\sim \exp (-Bs^{2-\gamma}),
\end{equation}
with $\gamma = 1-1/d\nu$, $d$ being the dimensionality of the system.
It is further suggested that the whole behavior of
$P(s)$ might well be described by
\begin{equation}
P(s) = As^\beta\exp (-Bs^{2-\gamma}),
\end{equation}
with $A$ and $B$ determined by the normalization condition,
and $\beta$ by the symmetry of the system.
Setting $\nu \approx 1.4$ \cite{MK},
we can estimate $\gamma$ at the three dimensional
Anderson transition to be 0.76, and the exponent of the
large $s$ behavior of ln$P(s)$ becomes
1.24.  The overall behavior of $P(s)$ has been claimed to be
well fitted to \cite{evan,HS}
\begin{equation}
A  s \exp (-Bs^{1.24}).
\end{equation}
It is further claimed that the same $P(s)$ has been obtained
in 3D disordered system without time reversal
symmetry \cite{HS2} and that it is consistent with the recent study on the
critical exponent in magnetic fields \cite{HKO}.

The linear rise of $P(s)$ even without time reversal symmetry
is controversial, since it is believed that the behavior of
$P(s)$ in the small $s$ region is determined by the symmetry of
the system, and in the absence of time reversal symmetry,
$P(s)$ should rise as $\sim s^2$.
The large $s$ behavior of $P(s)$ is recently questioned
by Zharekeshev and Kramer \cite{ZK}, who have performed the large size
diagonalization and asserted Poissonian behavior
$P(s)\propto \exp(-\kappa s)$
in large $s$ limit, with $\kappa < 1$.

Thus the existence of the critical level statistics at the Anderson
transition is well established, but it is fair to say that
the functional form of $P(s)$ is still controversial.

In this confusing situation, it is very important to study the
critical level statistics in other universality classes in the different
dimensionality where the extensive studies on the Anderson transition
have been performed.
One of the examples to show the Anderson transition is the
quantum Hall (QH) regime, where the Anderson transition takes place
at the Landau band center with the divergence of the localization length
as $1/|E|^\nu$, $E$ being the energy measured from the
Landau band center and $\nu\approx 2.4$ \cite{AA,HK}.
Independently of the above mentioned work on 3D transition,
the level statistics in the quantum Hall regime has been studied
by the present authors \cite{OKSOK,OO}.
It has been shown that the level statistics obeys the scaling behavior,
and the size independent level distribution
function $P(s)$ has been obtained at the Landau band center.
Another example is the two dimensional (2D) system with spin-orbit
interactions which derive the localized states to the extended ones
due to the anti-localization effect.

In this paper, we discuss the critical level statistics in two
dimensional
systems, especially $P(s)$,
at the Anderson transition in QH regime (unitary case)
as well as in the presence of strong spin-orbit scattering (symplectic case).
We have obtained the
energy spectrum by diagonalizing the Hamiltonian with randomness
for different systems and different sizes.
The energy eigenvalues are then unfolded in order to make the average
spacing unity \cite{OKSOK,OO}.
{}From the whole spectrum we take out levels in a central region with a width
of one tenth of the total width, and discuss their statistics.
The number of samples for each system size is chosen so that the
total number of levels for which the statistics are considered should not be
less than 25,000 in the case of QH regime and 50,000 in the symplectic case,
respectively.

First we discuss the QH case.
The disordered potential is assumed to be $\delta$-correlated.
Here we have adopted the random matrix Hamitonian
which describes the energy spectrum in the QH regime
\cite{HK,OKSOK}.
The area of the sample is varied from
200 to 800 in units of $2\pi \ell^2$ with $\ell$ the magnetic length.
In Fig. 1, the $P(s)$ at the Laudau band center
is shown in both linear (Fig. 1(a)) and logarithmic scales (Fig. 1(b)).
Different marks represent different system sizes.
These figures indicate that the critical distribution is size independent,
which is consistent with the scaling argument \cite{OO}.

Now we apply (2) using the value $\nu \approx 2.4$ which results in
$\gamma = 0.79$.
The result is shown by the broken lines.
We see considerable deviation from the numerical data.
Regarding $\gamma$ as a fitting parameter, very good agreement is
obtained with $\gamma = 0.35$, but this , together with $\gamma =1-1/d\nu$,
gives absurd value of $\nu (=0.77)$ that violates the
inequality $\nu > 2/d$ \cite{chay}.
It should be noted that
the small $s$ behavior of $P(s)$ is proportional to $s^2$, consistent
with the symmetry argument.

A similar difficulty also appears in the analysis of $P(s)$
at the Anderson transition in the presence of spin-orbit
interaction.  Using the Ando model \cite{ando} and estimating the critical
disorder by the finite size scaling method \cite{MK,ando,FABHRSWW},
we have obtained the critical level statistics.
In the numerical simulation, the ratio between the spin flip transfer
$V_2$ and the total transfer $V = \sqrt{V_1^2 + V_2^2}$
($V_1$ the non-spin-flip
transfer) is set to 0.5, and the critical disorder $W_{\rm c}$ is found to be
5.75$V$ and 6.7$V$ for box distribution and Gaussian distribution of
diagonal disorder, respectively.
In the case of the box distribution the parameter $W$ describing the
strength of disorder is defined as the width of the box, while in the
Gaussian distribution it is defined as $2\sqrt{3}$ times the
r.m.s. of the fluctuation of the site energy.
The factor $2\sqrt{3}$ is introduced so that both distributions may
have the same variance for the same value of $W$.
The linear system size is 8, 12, 16 and 20 in units of the lattice constant.

The results can be plotted in a similar way as in Fig. 1.
In Fig. 2, $P(s)$ at the Anderson transition  in the presence
of strong spin-orbit scatterings is plotted in the logarithmic scale.
It shows that the critical distribution is independent of the
system size and the type of disorder.
The behavior of the $P(s)$ is similar to the results
obtained recently by Schweitzer and Zharekeshev \cite{SZ}
who have diagonalized very large systems in the case of box type
potential distribution.
The critical distribution function $P(s)$ cannot be fitted to the theoretically
suggested form $P(s)\propto s^4\exp (-Bs^{2-\gamma})$ with $\gamma$ = 0.82
(broken line)
which corresponds to the critical exponent
$\nu = 2.8$ \cite{FABHRSWW}.
The best fit in the linear scale is obtained by putting
$\gamma = 0.35$.
Though this value of $\gamma$ does not reproduce the critical
exponent $\nu$,
good agreement in the small $s$ region supports $s^4$-rise of $P(s)$.

In summary, we have obtained the level spacing distribution function
in 2D quantum Hall regime as well as in the presence of the spin-orbit
interaction.
The level spacing functions $P(s)$'s at the
Anderson transition are shown to be independent of the system
size or of the type of the potential distributions.
In the QH case, $P(s)$ at the Anderson transition
rises as $s^2$ while it rises as $s^4$ in the
symplectic case.

The numerically obtained data can not be fitted to the theoretically
proposed functional form  (2) with $\gamma = 1-1/d\nu$.
This should not be taken seriously,
since the argument of ref. \cite{krav}
is valid only in the large $s$ region, and the overall functional
form, though very successful in 3D case,
is only a conjecture.
The numerical data in the large $s$ region qualitatively support
the asymptotic behavior predicted by them.
{}From the practical point of view, however, it is very difficult
to determine quantatively the behavior of $P(s)$ in this region,
since the number of data corresponding to large spacing is extremely
small.

In 3D, $P(s)$ at the critical points has been reported to
rise linearly in $s$ irrespectively of the symmetry of the system \cite{HS2}.
In ref. \cite{HS2}, the time reversal symmetry is broken by gauge fields
which can be absorbed into boundary conditions, and the system involves
no real magnetic field.
A study of more realistic systems will be necessary to clarify the behavior
of the critical statistics in 3D unitary ensemble.

The authors would like to thank Professor V.E. Kravtsov and Dr.
T. Kawarabayashi for fruitful
discussions.

\def\pr{Phys. Rev. }
\def\prl{Phys. Rev. Lett. }
\def\jpsj{J. Phys. Soc. Jpn. }
\def\ssc{Solid State Commun. }

\vfill\eject

{\bf Figure Caption}
\begin{description}
\item[Fig. 1] $P(s)$ in the quantum Hall regime in the linear (a) and
logarithmic
(b) scales. The circles ($\circ$), squares ($\Box$), diamonds ($\diamond$), and
triangles ($\triangle$) correspond to
the system area 200, 400, 600, and 800 in units of $2\pi\ell^2$.
 It can be very well fitted to the formula $As^2\exp(-Bs^{1.65})$
 indicated by the solid line.
 Applying (2) with $\gamma = 0.79$, we end up with poor fitting (broken line).

\item[Fig. 2] $P(s)$ in the presence of strong spin-orbit scatterings
 in the logarithmic scale.
 The circles ($\circ$), squares ($\Box$), diamonds ($\diamond$), and
triangles ($\triangle$) correspond to
the system size 8, 12, 16 and 20 in units of the lattice constant with
the box type potential distribution, while the crosses ($\times$)
and pluses ($+$) to the system size 16 and 20 with the Gaussian
type potential distribution.
The application of (2) with $\nu = 2.8$ and $\gamma = 1-1/d\nu$ is indicated
by the broken line, which shows considerable deviation in the small
$s$ region.
The best fit in the linear scale is obtained by setting $\gamma = 0.35$
(solid line).

\end{description}

\end{document}